\documentclass[aps,twocolumn,prd,preprintnumbers,amsmath,amssymb,superscriptaddress,nofootinbib]{revtex4-2}
\pdfoutput=1

\usepackage[export]{adjustbox}

 \usepackage{color}
 \usepackage{epsfig}
 \usepackage{ulem}

\definecolor{White}{rgb}{1,1,1}
\definecolor{Red}{rgb}{1,0.1,0}
\definecolor{LightYellow}{rgb}{1,1,.875}
\definecolor{SteelBlue}{rgb}{.273,.508,.703}
\definecolor{navy}{rgb}{0,0,.5}
\definecolor{LightCyan}{rgb}{.875,1,1}
\definecolor{DarkRed}{rgb}{.543,0,0}
\definecolor{HotPink}{rgb}{1,.41,.70}
\definecolor{ForestGreen}{rgb}{.13,.54,.13}
\definecolor{OliveDrab}{rgb}{.42,.55,.14}
\definecolor{MediumBlue}{rgb}{0,0,.80}
\definecolor{RoyalBlue}{rgb}{.25,.41,.88}
\definecolor{DeepSkyBlue}{rgb}{0,.746,1}
\definecolor{Brown}{rgb}{0.545,0.271,0.074}
\definecolor{Purple}{rgb}{0.637,0.285,0.641}

\usepackage{amsmath}
\DeclareMathOperator{\Tr}{Tr}

\begin{document}

\preprint{CTPU-PTC-21-36,  PNUTP-21-A15} 
 
\title{Electroweak baryogenesis by axionlike dark matter}

\author{Sang Hui Im}
\email{imsanghui@ibs.re.kr}
\affiliation{Center for Theoretical Physics of the Universe,   Institute for Basic Science, Daejeon 34126,   Korea }
\author{Kwang Sik Jeong}
\email{ksjeong@pusan.ac.kr}
\author{Yeseong Lee}
\email{yslee622@pusan.ac.kr}
\affiliation{Department of Physics,   Pusan National University,   Busan 46241,   Korea}

\begin{abstract}

We show that an axionlike particle (ALP)  naturally implements spontaneous electroweak baryogenesis 
through a cosmic evolution strongly tied to the electroweak phase transition (EWPT)  if
it feebly couples to the Higgs field   while giving a small contribution to the Higgs boson mass. 
The observed baryon asymmetry can be generated successfully if the ALP couples strongly enough 
to the electroweak anomaly.
Also interesting is that the ALP contributes to dark matter,  and 
its coupling to  a hidden gauge sector makes the relic abundance insensitive to the cosmic history
before the EWPT. 
The ALP explains both the baryon asymmetry and dark matter
in a wide range of the couplings owing to the friction induced by the hidden gauge sector.
To be compatible with cosmological and astrophysical observations, 
the ALP should have a mass  in the range between  about $0.01$ and $30$ eV,  
and is further required to  be photophobic if its coupling  to the electroweak anomaly 
is strong,    constraining the field content and charge assignment of the UV completion.

\end{abstract}  

\maketitle

\section{Introduction}

The observed matter-antimatter asymmetry and the cosmological constraints on dark matter
strongly point to the existence of new physics beyond the Standard Model (SM).   
The Higgs boson may provide a hint for the nature of dark matter and the origin of baryon asymmetry 
because it is sensitive to the details of UV physics and acts as a switch to decouple electroweak 
sphalerons that provide baryon number violation within the SM.   
The LHC experiments have found no definitive evidence of new physics 
at the TeV scale,  and showed only small deviations of the Higgs properties from their SM predictions. 
In addition,  although  had been long favored,
weakly interacting massive particles (WIMPs) face increasingly tight constraints. 
These results would indicate that a light particle feebly coupled to the SM resolves the puzzles
of the SM.

In this paper,   we show that an axionlike particle (ALP) coupled to the Higgs field can 
account for both the baryon asymmetry and dark matter of the universe.
The ALP feebly couples to the Higgs quadratic term with a coupling suppressed by a large decay
constant $F$,   such that its contribution to the Higgs boson mass is tiny. 
Then,  because it undergoes 
a cosmic evolution strongly tied to the electroweak phase transition (EWPT),  
which is a crossover as in the SM,
the ALP implements spontaneous electroweak 
baryogenesis~\cite{Cohen:1987vi,Cohen:1988kt,Cohen:1990py,Cohen:1991iu,Nelson:1991ab,DeSimone:2016ofp} 
via its coupling to the electroweak anomaly suppressed by a scale $f$.
Sufficient baryon asymmetry is generated for $F$ much larger than $f$,  and such a  large
hierarchy between the ALP couplings can naturally be achieved by the clockwork 
mechanism~\cite{Choi:2015fiu,Kaplan:2015fuy}.
The ALP also contributes to the dark matter of the Universe,  and the observed dark matter
abundance is obtained
if it couples to a hidden gauge sector to get thermal friction. 
The additional friction quickly fixes the ALP at the potential minimum well before the EWPT.
As a result,    the ALP relic abundance  becomes  insensitive to the cosmic history before the EWPT,
and is determined by the secondary oscillation driven by the EWPT. 
Interestingly,  the baryon asymmetry and dark matter are simultaneously explained by the ALP
in a wide range of $F$ and $f$ depending on how large friction is induced from 
the hidden gauge sector.

As coupled to the Higgs and electroweak gauge sectors,  the ALP is subject to 
various experimental constraints,  mainly to those associated with its longevity 
and astrophysical effects. 
The ALP mixes with the Higgs boson after electroweak symmetry breaking and consequently,
couples to other SM particles as well.
The ALP-Higgs mixing  should be small enough to make the ALP cosmologically stable and to be consistent
with astrophysical observations,  imposing stringent constraints on its mass depending on $F$.
Further,  in order to realize spontaneous baryogenesis,    the ALP should overcome 
the Hubble and thermal friction at temperatures above the EWPT.
This puts a lower bound on the ALP mass for $F$ and $f$ fixed to explain both the baryon 
asymmetry and dark matter.
It should be noted that the longevity and astrophysical constraints require 
the ALP to be photophobic~\cite{Craig:2018kne} 
for $f$ below about $10^7$~GeV since otherwise the coupling to the electroweak anomaly 
would make it too strongly couple to photons.    
The ALP turns out to have a mass in the range between about $0.01$ and  $30$~eV if it
simultaneously solves the puzzles of matter-antimatter asymmetry  and dark matter.

This paper is organized as follows. 
In Sec.~\ref{sec:evolution} we describe how the ALP coupled to the Higgs field evolves
as the Universe expands at temperatures above and at the EWPT,  
and how the evolution is affected by thermal friction.  
In Sec.~\ref{sec:Cogenesis} we discuss how to implement spontaneous electroweak
baryogenesis   and then show that the ALP can successfully generate both the baryon asymmetry 
and dark matter.
The experimental constraints on the ALP are discussed in Sec.~\ref{sec:Constraint}.
Section \ref{sec:conclusions} is the conclusions.

\section{ALP evolution}
\label{sec:evolution}

Let us begin with describing the cosmological evolution of an ALP coupled to the Higgs field.
Giving a small contribution to the Higgs boson mass via a coupling to the Higgs quadratic term,
the ALP can follow an evolution strongly tied to the EWPT.   
To be concrete,  we consider the following scalar potential:
\begin{equation}
V = \lambda (H^\dagger H)^2 
+ \mu^2_H(\phi) H^\dagger H+ V_0(\phi)  + \Delta V_{\rm TH},
\end{equation}
with the ALP dependent terms given by  
\begin{eqnarray}
\mu^2_H &=& -\mu^2 - M^2 \cos\left( \frac{\phi}{F} - \alpha \right),
\nonumber \\
V_0 &=& -\Lambda^4 \cos\left(  \frac{\phi}{F}   \right),
\end{eqnarray} 
which are presumably induced by some nonperturbative effects
explicitly breaking U$(1)_\phi$ symmetry of which the ALP is a 
pseudo Nambu-Goldstone boson.\footnote{
As a simple UV completion,  one can consider heavy vectorlike lepton doublets $L+L^c$
and singlets $N+N^c$,   which are charged under a hidden confining gauge group~\cite{Graham:2015cka}.
The gauge-invariant interactions,  $m_L LL^c + m_N NN^c + y HL N^c + y^\prime H^\dagger L^c N$,
then generate
$M^2 =  yy^\prime \Lambda^3_{\rm hid}/m_L$
at energy scales below $\Lambda_{\rm hid}$,
if the hidden confining scale $\Lambda_{\rm hid}$ lies in the range between $m_N$ and $m_L$
with $m_N\ll m_L$. 
}
Here,  $\Delta V_{\rm TH}$ includes thermal corrections,  and is dominated by
those from the SM particles at temperature $T\ll F$.
We are interested in the case where the ALP contribution to the Higgs boson mass
is negligibly small;  i.e.,   the case with 
\begin{equation}
M  \ll  v \equiv \frac{\mu}{\sqrt \lambda},   
\end{equation}
for which the EWPT occurs smoothly,   almost in the same way as in the SM,
but nonetheless it can significantly affect the ALP evolution.\footnote{
If $M$ is not much smaller than $\mu$,  the ALP evolution can
trigger the EWPT,  leading to interesting consequences,  such as
the cosmological relaxation of the electroweak scale~\cite{Graham:2015cka},  
the first-order EWPT required for electroweak baryogenesis~\cite{Jeong:2018ucz,Jeong:2018jqe},
spontaneous baryogenesis~\cite{Abel:2018fqg,Gupta:2019ueh},
and a solution to 
the dark matter problem~\cite{Abel:2018fqg,Gupta:2019ueh,Fonseca:2018kqf,Banerjee:2018xmn,Im:2019iwd}.
} 
Here,  $v\simeq 246$~GeV is  the Higgs vacuum expectation value at $T=0$. 
Further,   we take
\begin{equation}
\epsilon \equiv  \frac{M^2 v^2}{2 \Lambda^4}\sin\alpha \ll  1,
\end{equation}
as would be naturally the case because $\Lambda^4$ receives a radiative contribution 
 proportional
to $M^2$ from a closed Higgs loop,  up to a constant phase. 
The ALP mass is thus given by
\begin{equation}
m_\phi  \simeq \frac{\Lambda^2}{F},
\end{equation}
at $T=0$.

As in the SM, 
the EWPT occurs through a crossover transition,  and the Higgs background field value evolves 
according to\footnote{
The Higgs scalar potential receives  thermal corrections~\cite{Anderson:1991zb},
\begin{equation}
\Delta V_{\rm TH} \simeq
\frac{2 m^2_W + m^2_Z + 2 m^2_t}{8v^2} T^2 h^2
- \frac{2m^3_W + m^3_Z}{6\pi v^3} T h^3,
\nonumber 
\end{equation}
where $m_i$ is the mass of the corresponding particle.
The relation (\ref{higgs}) is obtained neglecting small corrections other than 
the quadratic term in $\Delta V_{\rm TH}$.
Numerical lattice simulations reveal a more precise relation for $h(T)$
at temperatures between $140$ and $170$~GeV~\cite{DOnofrio:2014rug,DOnofrio:2015gop}.
As we will see,   baryon asymmetry in our scenario is approximately proportional 
to $d h/d T$  at the temperature at which electroweak sphalerons become
effectively inactive.  
The baryon asymmetry obtained using (\ref{higgs}) is slightly smaller  than the one
estimated using the relation from  the lattice simulations.
}
\begin{equation}
\label{higgs}
h^2 \simeq v^2 \left( 1 - \frac{T^2}{T^2_c} \right),
\end{equation}
at $T$ below the critical temperature $T_c\simeq 153$~GeV,
where $h=\sqrt 2 |H^0|$ is the neutral Higgs scalar.  
The minimum of the potential appears along the ALP direction
at $\phi=0$ at $T$ above $T_c$,  but it shifts as the Universe cools down below $T_c$
as is obvious from the ALP potential approximately given by   
\begin{equation}
\Delta V \simeq  
- \Lambda^4 \cos\left(
\frac{\phi}{F} 
- \epsilon \frac{h^2}{v^2} 
\right),
\end{equation} 
for $\epsilon \ll 1$. 
At $T$ below $T_c$, 
the equation of motion for the ALP reads  
\begin{equation}
\label{eom}
\frac{d^2}{dt^2}
\frac{ \phi}{F}
+ \frac{3}{2 t } 
\frac{d }{dt }\frac{  \phi}{F} + m^2_\phi 
\left\{ 
\frac{\phi}{F}
- \epsilon \left( 1 - \frac{T^2}{T^2_c} \right)
 \right\} = 0 ,
\end{equation}
around the potential minimum,
where we have used that the Hubble expansion rate is given by $1/(2t)$ during a
radiation dominated era.  
Let us define $x\equiv m_\phi t$.
The solution of the above equation of motion is then found to be 
\begin{equation}
\label{ALP-soln0}
\frac{\phi}{F} \simeq
a  \left( \frac{x_c}{x} \right)^{3/4}
\sin( x- x_c - b)
+ \epsilon \left( 1 - \frac{x_c}{x}  \right),
\end{equation}
for $x_c\gg 1$,  
i.e.,  if the ALP mass is much larger than the Hubble expansion rate at $T_c$.
Here,   $x_c$ is the value of $x$ at $T_c$,
and the constants $a$ and $b$ are fixed by the initial conditions.  
The above shows that the ALP motion is composed of two parts.    
One is simply the shift of the potential minimum caused by the EWPT,
and  the other is coherent oscillation about the minimum.
Note that the ALP gets a kick at the EWPT
as can be seen from  the fact that $a$ and $b$ are given by
\begin{equation}
a = -\frac{\epsilon}{x_c},  
\quad
b = 0,
\end{equation}
if $\phi=d\phi/d t=0$ at $x=x_c$.
This implies that the EWPT makes the ALP oscillate with an amplitude proportional to $\epsilon$
even in the case where the ALP settles down to the potential minimum at  a temperature above $T_c$,

On the other hand, 
the ALP can couple to SM and/or hidden sector gauge bosons via the associated 
anomaly  as is naturally expected from its axionic nature.
Such a coupling can provide additional friction to the ALP motion~\cite{McLerran:1990de},
fixing quickly the ALP at the potential minimum.  
We parametrize the additional friction as $\Upsilon_\phi$, 
whose effect is included by taking the replacement, 
\begin{equation}
\frac{3}{2t} \frac{d}{dt}\frac{\phi}{F}\, 
\to
\left(
\frac{3}{2t}
+ \Upsilon_\phi
\right)
\frac{d}{dt}\frac{\phi}{F},
\end{equation}
in the equation of motion (\ref{eom}). 
Let us consider additional friction satisfying
\begin{equation}
\Upsilon_\phi \propto T^3,
\end{equation}
which is the case when the ALP couples to a hidden non-Abelian gauge sector  
as will be seen in Sec.~\ref{sec:Cogenesis}.
Then,  at $T<T_c$,   the solution of the equation of motion is written 
\begin{equation}
\label{ALP-soln}
\frac{\phi}{F} = 
y_1 \left( \frac{x_c}{x} \right)^{3/4}
\sin( x- x_c - b)
+ \epsilon \left( 1 - \frac{x_c}{x}  
+ y_2 
\right),
\end{equation}
for $x_c\gg 1$, 
where $y_1$ and $y_2$ are a function of $x$. 
For the case with $\phi=d\phi/dt =0$ at $x=x_c$,  one finds that 
$b$ is given by $b= \arcsin(\kappa x^{-3/2}_c)$,  while $y_1$ and $y_2$ 
approximately
read  
\begin{equation}
y_1 \simeq -\frac{\epsilon}{x_c} e^{ \frac{\kappa}{\sqrt x} - \frac{\kappa}{\sqrt{x_c}} },
\quad
y_2 \simeq - \kappa \frac{x_c}{x^{7/2}}, 
\end{equation}
assuming $\kappa<x^{3/2}_c$.
Here,  $\kappa$ is defined by
$\kappa \equiv\sqrt{m_\phi t^3} \Upsilon_\phi$,
which can be treated as a constant during a radiation dominated era
if  $\Upsilon_\phi$ is proportional to $T^3$. 
It should be noted that the oscillation amplitude,  which is still proportional
to $\epsilon$,  can get exponentially suppressed depending on 
the value of $\kappa$.

\begin{figure}[t] 
\begin{center} 
 \begin{tabular}{c}
      \includegraphics[width=0.45 \textwidth]{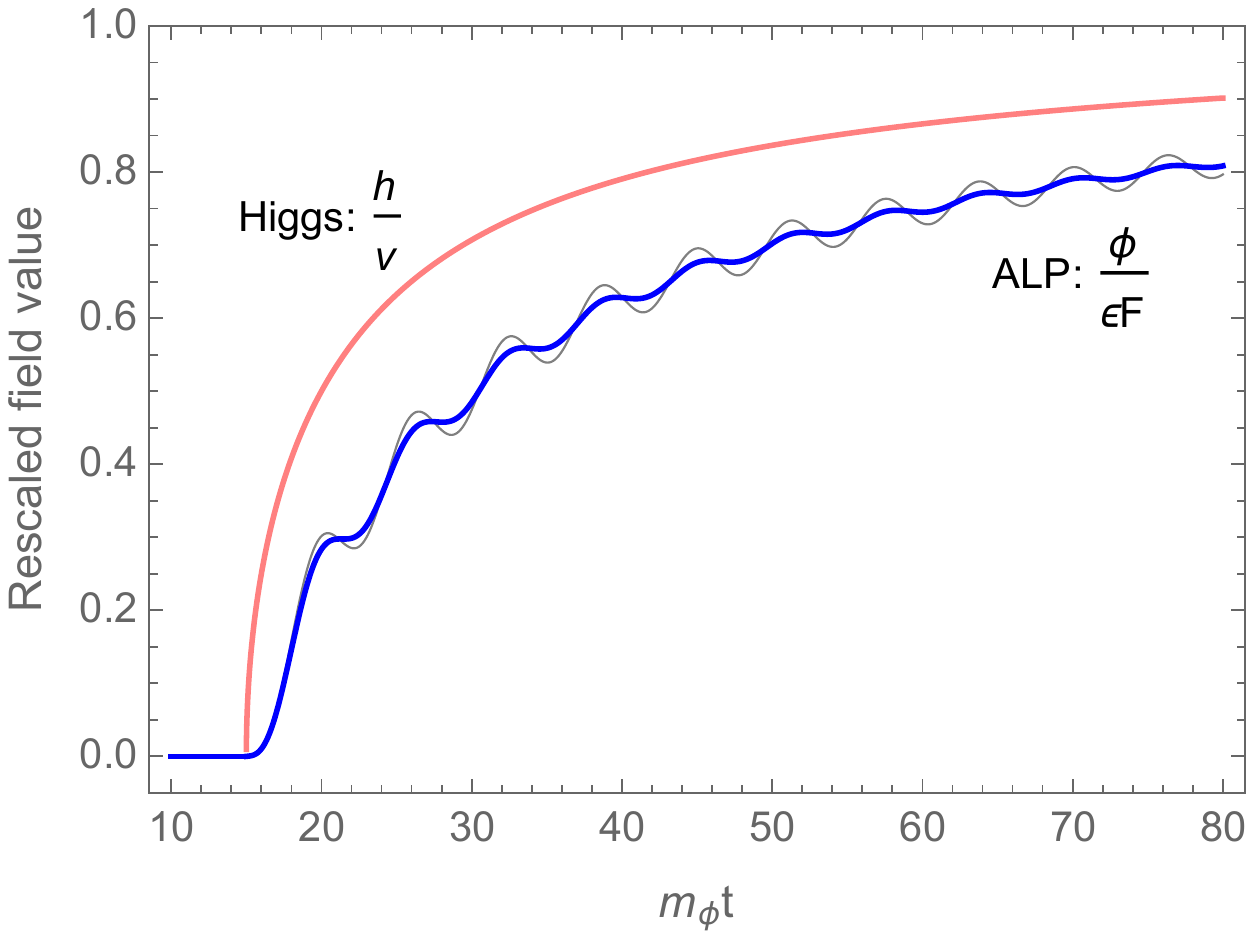} 
  \end{tabular} 
 \end{center}
 \caption{
Cosmic evolution of the Higgs and the ALP coupled to the Higgs field
for $m_\phi=1$~meV.
The pink curve is for the Higgs field,  $h(t)/v$,  while
the blue (gray) curve is for the ALP,  $\phi(t)/ \epsilon F$,
with $\kappa=10$ ($\kappa=0$).
Here,  we have assumed that the ALP is fixed at the potential minimum well before the EWPT.
}
 \label{fig:ALP-evolution} 
\end{figure}

Figure \ref{fig:ALP-evolution} illustrates how the ALP coupled to the Higgs field 
evolves while tied to the EWPT,   and how much it is affected by the additional friction.
The ALP oscillates around the potential minimum,  which is shifted 
during the EWPT. 
The oscillation amplitude decreases due to the expansion of the Universe as shown by
the gray curve,  and the damping is exponential in the presence of  
the additional friction proportional to $\kappa$ as shown by the blue  curve.
We have checked that the approximate analytic solution (\ref{ALP-soln})
is in precise agreement with the numerical results.

\section{Baryon and dark matter genesis}
\label{sec:Cogenesis}

The ALP couples to the Higgs sector while giving a small contribution
to the Higgs boson mass.  
As a result,   the potential minimum around which the ALP undergoes a damped oscillation
smoothly shifts as the Universe cools down below $T_c$. 
It is during the EWPT that such shift of the potential minimum occurs,  allowing the ALP
to implement spontaneous electroweak baryogenesis.
Furthermore,  the ALP coherent oscillation can contribute to dark matter.

As a source of $CP$ violation for baryogenesis,  we add
an ALP coupling to the electroweak anomaly,
\begin{equation}
\label{ALP-EW}
\Delta {\cal L} 
= \frac{1}{16\pi^2}\frac{\phi}{f}  \Tr  W^{\mu\nu} \tilde W_{\mu\nu},
\end{equation}
which can be induced,  for instance,  by the loops of heavy
leptons carrying a U$(1)_\phi$ charge.
Because the baryon and lepton numbers,   $B$ and $L$,  are anomalous in the SM
while their difference is not, 
the baryon number density evolves as
\begin{equation}
\label{BN-evolution}
\frac{d n_B}{d t} = N_g  \left\langle \frac{1}{16\pi^2} 
\Tr  W^{\mu\nu} \tilde W_{\mu\nu} \right\rangle,
\end{equation}
where $N_g=3$ counts the number of generations,  and
the thermal average of the Chern-Simons (CS) number density 
is determined by the rate at which a chemical potential drives topological transitions, 
which is related to the free diffusion rate for topological charge $ \Gamma_{\rm sph}$ 
by the factor $1/2T$~\cite{Moore:1996qs,Berghaus:2020ekh}, 
\begin{equation}
\label{CS-thermal}
\left\langle \frac{1}{16\pi^2} 
\Tr  W^{\mu\nu} \tilde W_{\mu\nu}, \right\rangle
= 
\frac{\Gamma_{\rm sph}}{2 T} \left(
\frac{d}{dt}\frac{\phi}{f}
- \frac{13}{2} \frac{n_B}{T^2}
\right),
\end{equation}
with $\Gamma_{\rm sph}$ being
the electroweak sphaleron transition rate per unit volume.
In the CS number density,   the term proportional to $d\phi/dt$ 
is the driving force,  while the other is the energy cost of producing CS number
against the chemical potential.
Here we have used that  
each sphaleron creates nine quarks and three leptons,  and thus the CS chemical potential
is given by $\mu_{\rm CS} = \sum \mu_q + \sum \mu_\ell =13n_B/(2T^2)$
with  $\mu_q$ and $\mu_\ell$ being the quark and lepton chemical potential,
respectively~\cite{Bochkarev:1987wf,Cline:2000nw}. 
The above shows that the time derivative of $\phi$ serves as
the chemical potential of baryon number,  generating baryon asymmetry 
via the electroweak anomaly.

When the friction decreases enough, 
the ALP starts coherent oscillations with an initial misalignment generally of the order of $F$.
However, 
if the cosmic expansion is the only friction,   the ALP relic abundance exceeds  the observed value.
To avoid it,  one may rely on the anomalous coupling (\ref{ALP-EW}) because it  affects the equation of motion 
for the ALP by inducing thermal friction given by   
\begin{equation}
\label{SM-friction}
\Upsilon_\phi|_{\rm SM} = \frac{\Gamma_{\rm sph}}{2 T f^2},
\end{equation}
as is obvious from the time-derivative term in (\ref{CS-thermal}).
For the ALP much lighter than the MeV scale,   which is the region of our interest,  
however, 
the baryon number density promptly approaches to a constant
within the time 
\begin{equation}
\Delta t \sim 
\frac{4}{13 N_g}\frac{T^3}{\Gamma_{\rm sph}}
\sim \frac{10^{5}}{ T},
\end{equation}
thereby canceling the thermal friction.
Here,  we have used that $\Gamma_{\rm sph} \sim 10^{-6} T^4$ in the
symmetric phase. 
Note also that the baryon asymmetry rapidly oscillates between negative and positive values,  
making it difficult to predict its final value.

To provide additional large friction,  we consider  a hidden non-Abelian gauge sector, 
such that 
classical global symmetries other than  U$(1)_\phi$,   if any,   are all anomaly free,
and spontaneous gauge symmetry breaking occurs at a temperature 
below $T_c$ but before the big bang nucleosynthesis.
In such a case, 
if the ALP couples to the hidden gauge sector through\footnote{
The ALP may couple to the hidden gauge sector with a periodicity scale $f^\prime$ instead of $f$.
However,   this  does not change our results because 
the baryon asymmetry produced
by the ALP evolution and the ALP contribution to dark matter depend on
the hidden gauge sector only through the combination 
$\zeta_\phi/f^2$ with $\zeta_\phi$ defined in (\ref{hid-friction}).
Our results remain the same under $f\to f^\prime$ and $\zeta_\phi \to \zeta_\phi (f^\prime/f)^2$. 
}
\begin{equation}
\frac{1}{16\pi^2} \frac{\phi}{f} \Tr W^{\prime   \mu\nu} \tilde W^{\prime }_{\mu \nu},
\end{equation}
the CS number density includes only the driving force $d\phi/dt$ acting on it,
\begin{equation}
\label{ALP-hidden}
\left\langle \frac{1}{16\pi^2} 
 \Tr W^{\prime   \mu\nu} \tilde W^{\prime }_{\mu \nu} \right\rangle
= \frac{\Gamma^\prime_{\rm sph}}{2T} \frac{d}{dt} \frac{\phi}{f},
\end{equation} 
and  a sphaleron event in the hidden sector has no energy cost of producing particle number
against a chemical potential.
Here,  $\Gamma^\prime_{\rm sph}$ is the hidden sector sphaleron rate. 
Thus,  
the coupling to the hidden gauge sector creates friction for the ALP,
\begin{equation}
\label{hid-friction}
\Upsilon_\phi|_{\rm hid} = \zeta_\phi \frac{T^3}{f^2}, 
\end{equation}
in the symmetric phase during which the sphaleron rate is proportional to $T^4$. 
Here,  $\zeta_\phi$ is determined by the hidden gauge coupling;
for instance,   it is of the order of $N^5_c \alpha^{\prime 5}$ 
for a gauge group SU$(N_c)$ with  $\alpha^\prime$ being the gauge fine structure constant.
The above friction  quickly suppresses the oscillation amplitude
according to (\ref{ALP-soln}) and fixes the ALP at the potential minimum 
before  the EWPT starts.\footnote{
The ALP dissipates its energy into the hidden sector plasma.
To avoid cosmological constraints,  we simply assume that  hidden sector particles eventually decay 
into the SM particles quickly enough,
for instance,  through renormalizable SM operators coupled to the gauge invariant quadratic term 
of the hidden sector Higgs field.
Note also that,   if the hidden sector plasma is colder than the SM one,  
$\zeta_\phi$ gets suppressed by the temperature ratio cubed.
}
As a result,     the ALP relic abundance becomes insensitive to the cosmic
history before the EWPT  and is determined solely by the secondary oscillation
induced by the kick at the EWPT.

The recent lattice simulation~\cite{DOnofrio:2014rug} shows how the
electroweak  sphaleron rate changes with
the temperature and the Higgs background field value.
The sphaleron rate reads $\Gamma_{\rm sph}  \approx 18\alpha^5_W T^4$ 
at high temperatures,   where  $h$ is close to zero,   while it is exponentially suppressed by the Boltzmann factor
$e^{-E_{\rm sph}/T}$ if $h$  is larger than about $0.5 T$. 
Here,   $E_{\rm sph}\simeq 4\pi h/g$ is the sphaleron energy.
This indicates that sphalerons become effectively inactive when the Universe cools down below  
the temperature,
\begin{equation}
T_{\rm cut} \approx 0.95 T_c,
\end{equation}  
which has been numerically estimated using
the approximated relation (\ref{higgs}) for the Higgs field value.
Note that the amounts of baryon and dark matter in our scenario rely on 
the ratio between $T_{\rm cut}$ and $T_c$ rather than their precise values. 
It is worth noting also that,   at    $T\gtrsim T_c$, 
the ALP can decay into the electroweak or hidden gauge bosons via the couplings 
(\ref{ALP-EW}) and (\ref{ALP-hidden}).
The decay rate is however negligibly small compared to the thermal dissipation rate
as long as the ALP mass is much smaller than the  electroweak scale. 

Before delving into the detailed analysis,  let us set the parameter region of our interest.
The ALP should be heavy enough to overcome the Hubble and thermal friction during
the EWPT to implement baryogenesis. 
There are also strong experimental constraints associated with  ALP-Higgs mixing 
as discussed in Sec.~\ref{sec:Constraint}.
These require  the ALP mass to be in the range,    
\begin{equation}
\label{ALP-mass}
0.01\,{\rm eV} \lesssim m_\phi \lesssim 30\,{\rm eV},
\end{equation}
depending on the value of $\epsilon F$.
In addition,   in order to generate sufficient baryon asymmetry,  
the ALP couplings should be 
\begin{equation}
f \ll  \epsilon  F.
\end{equation}
A large hierarchy between $f$ and $F$  can be achieved in a natural way via
the clockwork mechanism,
where collective rotations of multiple axions generate an exponential hierarchy
between the couplings of the lightest mode~\cite{Choi:2015fiu,Kaplan:2015fuy}.

Let us now examine how much baryon asymmetry can be produced. 
The electroweak sphaleron transitions are rapid in the symmetric phase,
violating the baryon and lepton numbers~\cite{Kuzmin:1985mm,Giudice:1993bb,Rubakov:1996vz}.   
In such circumstances, 
baryon asymmetry is generated according to (\ref{BN-evolution}),   
\begin{equation}
\label{nB}
\frac{d n_B}{d t} =
\frac{N_g}{2} \frac{\Gamma_{\rm sph}}{T} 
\frac{d}{d t} \frac{\phi}{f}
- \Gamma_B n_B,
\end{equation}
for the ALP coupled to the electroweak anomaly (\ref{ALP-EW}).
Here,  $\Gamma_B = (13N_g/4) \Gamma_{\rm sph}/T^3$
is the rate of sphaleron-induced relaxation of baryon asymmetry~\cite{Bochkarev:1987wf,Cline:2000nw}. 
Successful baryogenesis manifestly requires some way to overcome the washout effect.
One way,  which is known as electroweak baryogenesis,  is to produce baryon asymmetry at the electroweak scale 
just before the sphalerons are decoupled.
As monotonically varying  during the EWPT neglecting rapid oscillations,
the ALP coupled to the Higgs field can naturally realize such a scenario
even if the EWPT is not strongly first order.

\begin{figure}[t] 
\begin{center} 
 \begin{tabular}{c}
      \includegraphics[width=0.45 \textwidth]{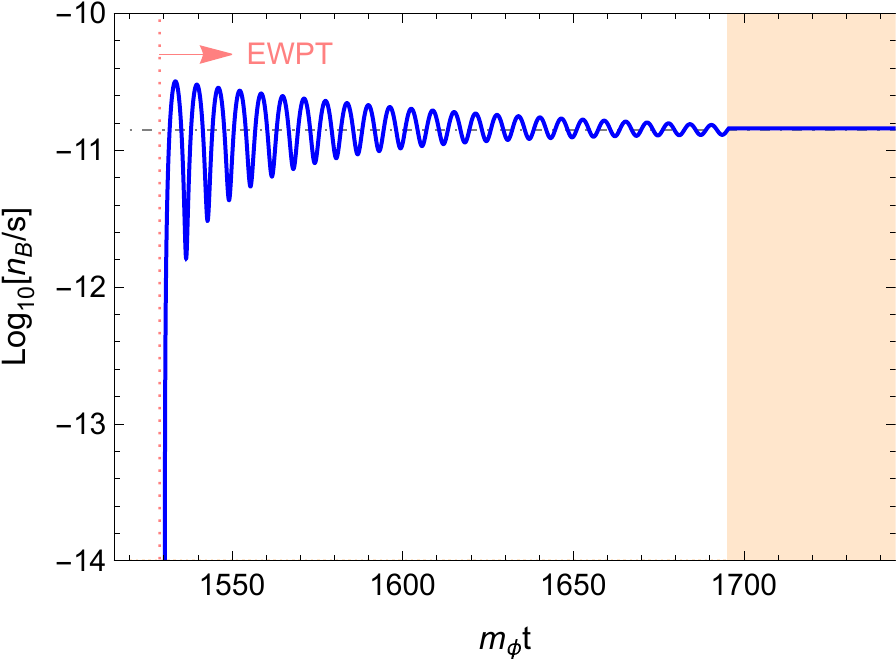} 
  \end{tabular} 
 \end{center}
 \caption{
Spontaneous baryogenesis driven by the ALP coupled to the Higgs field
in the case with $\zeta_\phi=10^{-4}$,
$m_\phi=0.1$~eV,  $f=10^7$,  and $\epsilon F=10^{14}$~GeV.
Baryon asymmetry is produced via rapid sphaleron processes during the ALP evolution
while balancing the production against washout.
When the Universe cools down below $T_{\rm cut}$,  which corresponds to
the shaded region, 
the baryon-to-entropy ratio  is  frozen to
around the value (\ref{nBs}) shown by the horizontal line.
}
 \label{fig:BAU} 
\end{figure}

On the quantitative level,  it is found that
sufficiently large $\epsilon F/f$ makes $n_B$ quickly approach to the value such that 
the two processes in (\ref{nB}) associated with the baryon production and washout  
balance each other. 
From (\ref{ALP-soln}) and (\ref{nB}),  one can thus approximately estimate 
the baryon-to-entropy ratio, 
\begin{equation}
\label{nBs}
\frac{n_B}{s} \simeq \left. \frac{n_B}{s}\right|_{T_{\rm cut}}
\approx
1.4 \times 10^{-18} \frac{\epsilon F}{f},
\end{equation} 
neglecting the oscillating piece which is rapidly damped away due to the thermal friction with
\begin{equation}
\kappa \simeq 2.1 \times 10^3
\left( \frac{\zeta_\phi}{10^{-4} } \right)
\left( \frac{m_\phi}{0.1{\rm eV}} \right)^{1/2}
\left( \frac{f}{10^7 {\rm GeV}} \right)^{-2}, 
\end{equation}
induced by the ALP coupling to the hidden gauge sector.
Here,   we have used that the baryon-to-entropy ratio  is frozen at $T_{\rm cut}$  
because sphalerons become effectively inactive,
stopping both the baryon production and washout processes.
The observed baryon asymmetry is therefore explained by the ALP with
\begin{equation}
\label{BAU-condition}
\frac{\epsilon F}{f} \approx
0.6 \times 10^8.
\end{equation}
Figure \ref{fig:BAU} shows how spontaneous baryogenesis is  induced by the ALP.
Here,   we have taken
$\zeta_\phi=10^{-4}$,   $m_\phi=0.1$~eV,  $f=10^7$~GeV,  and $\epsilon F=10^{14}$~GeV.   
Combined with rapid sphaleron processes,  the ALP evolution produces 
baryon asymmetry in such a way that the production balances with the washout
for  $\epsilon F$ much larger $f$.
We have numerically confirmed this feature  to hold.
The sphalerons become inactive below $T_{\rm cut}$,  fixing the baryon asymmetry
around the value estimated by (\ref{nBs}).

We continue to evaluate the ALP relic abundance. 
The ALP gets a kick at the EWPT,  and then undergoes highly damped oscillation 
due to the thermal friction induced by its coupling to the hidden gauge sector
until the universe cools down below the critical temperature,  $T_{\rm hid}$,
at which  the hidden gauge symmetry breaking occurs. 
Afterwards,   as subject only to the Hubble friction,  
the ALP evolves according to (\ref{ALP-soln0}),  implying that its coherent oscillations 
behave like cold dark matter.
From these facts,   the ALP relic density is  found roughly to be
\begin{equation}
\label{DM}
\Omega_\phi h^2
\approx
0.12 \left(
\frac{e^{-\beta} \epsilon F}{0.5 \times 10^{13}{\rm GeV} }
\right)^2,
\end{equation}
with the exponent $\beta$ given by
\begin{equation}
\beta = \frac{\kappa}{\sqrt{x_c}} - \frac{\kappa}{\sqrt{x_{\rm hid}} }
\simeq
54 
\left( \frac{\zeta_\phi}{10^{-4}} \right)
 \left(\frac{f}{10^7{\rm GeV}}\right)^{-2},
\end{equation}
where the last approximation holds for $T_{\rm hid}$ much below $T_c$.
The ALP can thus account for the dark matter relic density of the Universe if
\begin{equation} 
\label{DM-condition}
e^{-\beta}  \epsilon F \simeq 
0.5  \times 10^{13}\,{\rm GeV}.
\end{equation} 
assuming that the ALP is cosmologically stable.  
We shall discuss how to ensure  the longevity of the ALP 
for $f\ll \epsilon F$
in Sec.~\ref{sec:Constraint}.

\begin{figure}[t] 
\begin{center} 
 \begin{tabular}{c}
      \includegraphics[width=0.45 \textwidth]{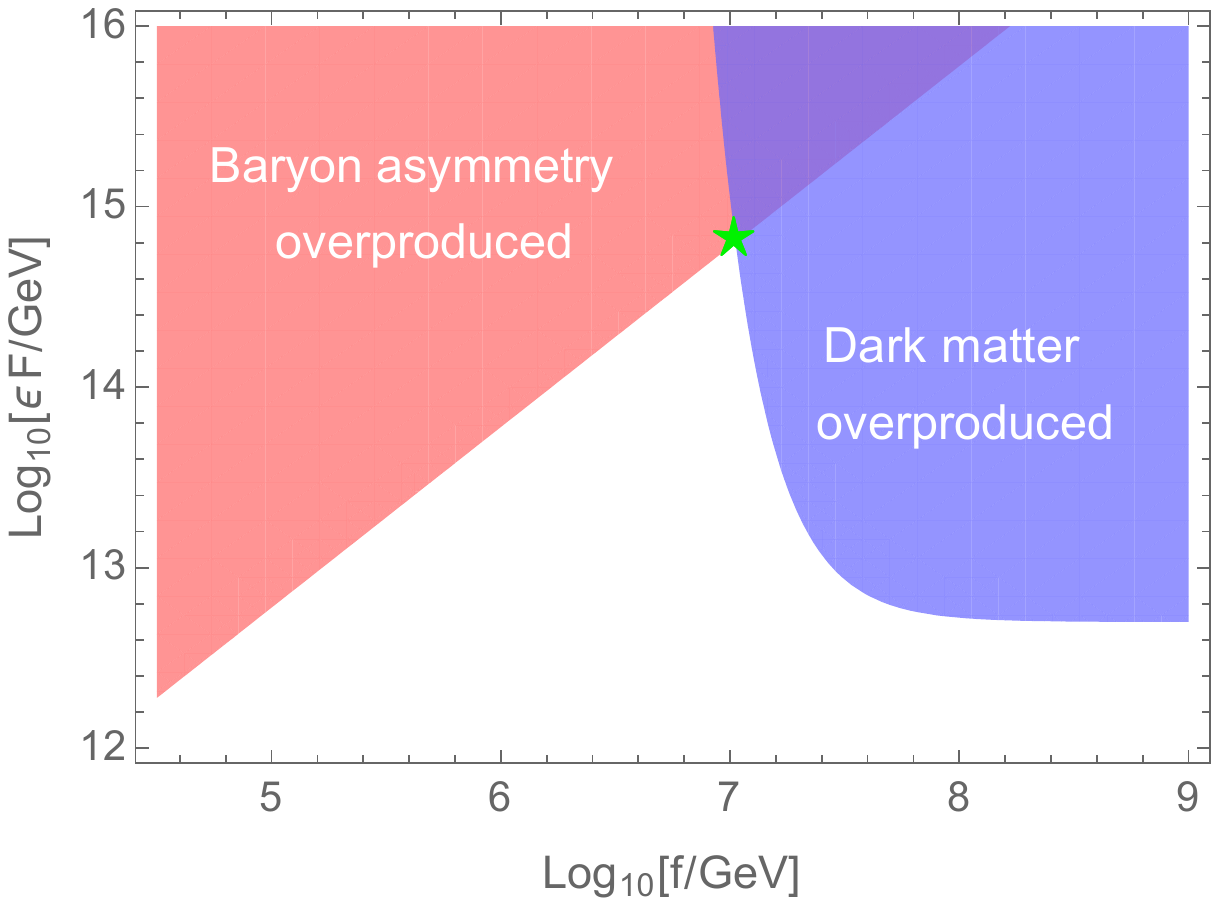} 
  \end{tabular} 
 \end{center}
 \caption{
Viable parameter region of the ALP couplings to produce
baryon asymmetry and dark matter for  $\zeta_\phi=10^{-5}$.
The red and blue shaded regions lead to overproduction of
the indicated component.   
The ALP coupled to the Higgs boson explains  both the baryon asymmetry and 
dark matter of the Universe
at the filled green star.
 }
 \label{fig:BAU-DM} 
\end{figure}

An intriguing feature of our scenario is that  the baryon asymmetry (\ref{nBs}) 
and the dark matter relic density (\ref{DM}) are determined only by  
$\epsilon F$,   $f$,  and $\zeta_\phi$.
Remarkably,   the observed baryon and dark matter densities of the Universe
are simultaneously explained by the ALP  if it couples to the Higgs and the electroweak
gauge bosons with  
\begin{eqnarray}
\label{cogenesis}
\epsilon F &\approx&
1.8\times 10^{15}\,{\rm GeV} 
\left(
\frac{\zeta_\phi}{10^{-4}}
\right)^{0.46},
\nonumber \\
f &\approx& 3 \times 10^7 \,{\rm GeV}
\left(
\frac{\zeta_\phi}{10^{-4}}
\right)^{0.46},
\end{eqnarray}
for $\zeta_\phi\ll 1$
and has a mass in the range between  about $0.01$ and $30$~eV to avoid the experimental constraints. 
Figure \ref{fig:BAU-DM} shows the viable region of $\epsilon F$ and $f$ for the ALP-driven cogenesis
of baryon and dark matter,
where we have taken $\zeta_\phi=10^{-5}$. 
The red and blue shaded regions lead to overproduction of baryon 
and dark matter,    respectively,
putting an upper bound on the ALP coupling to the Higgs field.
The cogenesis point explaining both the baryon asymmetry and dark matter 
is indicated by the filled green star.

\section{Experimental constraints}
\label{sec:Constraint}

In this section,   we examine experimental constraints on the ALP mass and couplings.
The ALP-Higgs mixing is strongly constrained  because
it allows the ALP to have various astrophysical effects. 
Furthermore,  its longevity should be ensured if the ALP is to constitute
a fraction or all of the dark matter we detect today.

The ALP interacts with the Higgs and electroweak gauge bosons with a coupling
suppressed by $F$ and $f$,  respectively,   with $f\ll \epsilon F$.
As coupled to the Higgs quadratic term,  the ALP mixes with the Higgs boson 
after electroweak symmetry breaking, 
and consequently,  it couples also to other SM particles.
The mixing angle is estimated by
\begin{equation}
\theta_{\rm mix} 
\simeq 0. 5\times 10^{-13}
\left( \frac{\epsilon F}{10^{13}{\rm GeV} } \right)
\left( \frac{m_\phi}{0.1{\rm eV}} \right)^2,
\end{equation}
which should be small enough to make the ALP cosmologically stable. 
If heavier than the MeV scale,  the ALP decays into charged leptons  that 
subsequently produce photons via
final state radiation~\cite{Essig:2013goa}
and inverse Compton scattering~\cite{Blanco:2018esa}.  
The mixing is then severely constrained by gamma ray observations,
requiring the ALP to be lighter than the electron.
If lighter than the electron,  the ALP decays  dominantly into a pair of photons with a lifetime given by
\begin{equation}
\tau_\phi \simeq 6\times 10^{35}\,{\rm sec}
\left( \frac{\epsilon F}{10^{13}{\rm GeV} } \right)^{-4}
\left( \frac{m_\phi}{0.1{\rm eV} }\right)^{-7}.
\end{equation}
For $m_\phi$ above $10$~eV,  the lifetime should be longer than about $10^{25}$~sec 
to avoid the constraints from the extragalactic background light (EBL) and 
the ionization of primordial hydrogen caused by the produced photons
if the ALP constitutes a major fraction of dark matter~\cite{Cadamuro:2011fd}.
The ALP couples to the electron through the mixing and thus,   
 is subject to a more stringent constraint arising from the stellar cooling
bound if its mass is below keV~\cite{Hardy:2016kme}.
This requires $\theta_{\rm mix}$ to be less than $0.3\times 10^{-9}$ in magnitude.
The mixing is further constrained if the ALP is lighter than an eV
because it can then mediate a long-range force through the Yukawa couplings to the SM
particles~\cite{Kapner:2006si,Schlamminger:2007ht,Berge:2017ovy}

\begin{figure}[t] 
\begin{center} 
 \begin{tabular}{c}
      \includegraphics[width=0.45 \textwidth]{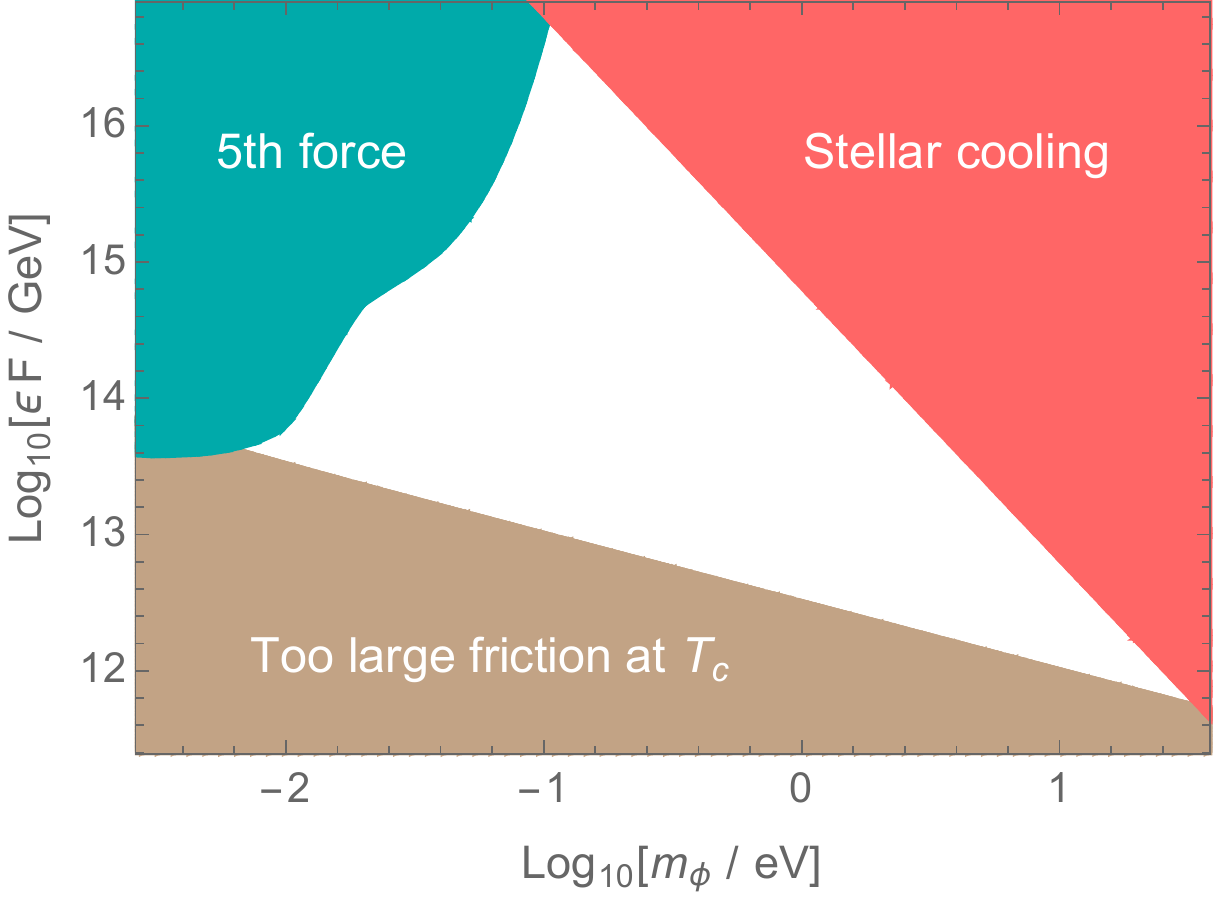} 
  \end{tabular} 
 \end{center}
 \caption{
Constraints on the ALP mass and ALP-Higgs coupling
for the ALP satisfying the condition  (\ref{cogenesis})
to explain  both  the baryon asymmetry 
and dark matter,  which fixes the coupling to the electroweak anomaly to
be $f\simeq 1.7\times 10^{-8} \epsilon F$.
The shaded regions are excluded by the indicated constraints,  respectively. 
}
 \label{fig:constraints} 
\end{figure}

The constraints on the ALP mass and coupling are summarized in Fig.~\ref{fig:constraints}. 
The dark cyan and red shaded regions are excluded by the constraints  from the long-range 
fifth force and the stellar cooling,  respectively.
Further,  to implement spontaneous baryogenesis,  the ALP mass should be larger than the 
Hubble and thermal friction at $T_c$.
Here,   the thermal friction should include
the contribution from the electroweak gauge sector 
because its effect cancels only after the ALP starts to move,  i.e., 
\begin{equation}
\Upsilon_\phi  \simeq 0.4\times 10^{-6}\frac{T^3_c}{f^2}  + \zeta_\phi \frac{T^3_c}{f^2}.
\end{equation} 
For the ALP couplings satisfying  the cogenesis condition (\ref{cogenesis}), 
the thermal friction is always larger than the Hubble friction at $T_c$,
and the lower bound on the ALP mass is put by the electroweak gauge contribution in the region 
compatible with the fifth force constraint, 
\begin{equation}
10^{-4}\,{\rm eV}
\left\{
1+ \left( \frac{3.2\times 10^{14}{\rm GeV}}{\epsilon F} \right)^2 \right\}
< m_\phi,
\end{equation}
excluding the brown shaded region. 
The constraints discussed so far indicate that 
the ALP should have 
\begin{equation}
0.01 \lesssim \frac{m_\phi}{\rm eV} \lesssim 30,
\quad
10^{12} \lesssim \frac{\epsilon F}{\rm GeV}
\lesssim 10^{17},
\end{equation} 
with $f\simeq 1.7\times 10^{-8} \epsilon F$, 
if it is to be responsible both for the baryon asymmetry and dark matter.

The ALP coupling to the electroweak anomaly (\ref{ALP-EW}),  which is  essential in implementing
the cogenesis of baryon and dark matter,   induces a coupling to photons,  
\begin{equation}
c_{\rm EM} \frac{\alpha_{\rm EM}}{4\pi} \frac{\phi}{f}
F^{\mu\nu} \tilde F_{\mu\nu},
\end{equation}
after electroweak symmetry breaking. 
For the ALP with mass in the range (\ref{ALP-mass}),  the strongest constraint on the ALP-photon coupling 
comes from the cooling rate of horizontal branch stars~\cite{Ayala:2014pea},
\begin{equation}
\label{ALP-photon}
 \frac{1}{c_{\rm EM}} 
f  > 3.5 \times 10^7\,{\rm GeV},
\end{equation}
for $m_\phi$ below $10$~keV.
The CERN Axion Solar Telescope (CAST) also puts a similar bound for $m_\phi $ below $0.02$~eV,  
and a slightly weaker bound suppressed by a factor of about $3$ for $m_\phi$ between 
$0.02$ and $1$~eV~\cite{CAST:2017uph}.  
For $m_\phi>10$~eV,  $f/c_{\rm EM}$ should be larger than about $10^{10}$~GeV
because the astrophysical constraints require the ALP to live longer than $10^{25}$~sec
as discussed above.
The constraints on $f$ are evaded if  $c_{\rm EM}$ is suppressed much below unity. 
This is indeed the case when the coupling (\ref{ALP-EW}) is generated by the loops of heavy leptons 
charged under SU$(2)_L$ and/or U$(1)_Y$ with the U$(1)_\phi$ charge assignment such that
the chiral anomaly U$(1)_\phi\times$U$(1)^2_{\rm EM}$ vanishes,  
i.e.,  if the ALP is photophobic~\cite{Craig:2018kne}. 
Then,  
the coupling to the electroweak gauge bosons induces only a tiny coupling  at one and two loops,
\begin{equation}
c_{\rm EM} = \frac{k_W }{16\pi^2} \frac{m^2_\phi}{m^2_W}
+ \frac{k_F}{(16\pi^2)^2} \frac{m^2_\phi}{m^2_F}, 
\end{equation}
for the order unity constants $k_W$ and $k_F$,
because the U$(1)_\phi$ symmetry is explicitly broken by the ALP mass. 
Here,  $m_W$ is the $W$-boson mass,  and $m_F$ is the mass of U$(1)_\phi$ charged leptons. 
Note that such photophobic nature does not require fine-tuning of the involved model parameters
but is a result of a proper charge assignment of heavy leptons.

On the other hand,    one can consider a model where 
the ALP couples also to the QCD anomaly,
i.e.,  a model where there are quarks or heavy colored fermions carrying 
a U$(1)_\phi$ charge such that the anomaly
U$(1)_\phi\times SU(3)^2_c$ does not vanish.
The coupling to the QCD,  if exists,  should be suppressed by a scale greater than about 
$10^9$~GeV  
in order to satisfy various astrophysical constraints~\cite{Choi:2020rgn}.
In addition,  though will not be explored here,   it can slightly modify 
the ALP relic density  
because the scalar potential of the ALP
receives an additional contribution during the QCD phase transition.

Finally,   we briefly  discuss the experimental constraint  from the electric dipole moment (EDM)
of the electron.  
In the presence of mixing with the Higgs boson, 
the ALP violates $CP$ symmetry through its coupling to the electroweak anomaly.
As a result,  the electron EDM is radiatively generated~\cite{Choi:2016luu,Flacke:2016szy},  
\begin{equation}
d_e \simeq \frac{8e^3}{(16\pi^2)^2}\frac{m_e}{v}
\frac{c_{\rm EM} \theta_{\rm mix}}{f}
\ln\left( \frac{m_h}{m_\phi} \right),
\end{equation}
where $m_h$ and $m_e$ are the mass of the Higgs boson and the electron,
respectively. 
The above contribution  is however much below the experimental bound~\cite{Andreev:2018ayy}
because the ALP-Higgs mixing is tiny in the parameter space of our interest,
and it is further suppressed to be negligible  if the ALP is photophobic.
\\

\section{Conclusions}
\label{sec:conclusions}

We have shown that an ALP coupled to the Higgs field can provide the observed dark matter 
abundance,   and further relate it with the matter-antimatter asymmetry of the Universe.
As strongly tied to the EWPT,  the cosmic evolution of the ALP naturally provides $CPT$ violation suitable
for spontaneous electroweak baryogenesis  
in the presence of a strong enough coupling to the electroweak anomaly. 
Another intriguing feature  is that  the ALP relic abundance is determined
by the secondary coherent  oscillation driven by the EWPT,
insensitively to  the cosmic history before the EWPT,   if the ALP couples to a hidden gauge
sector to get additional friction.    
The ALP is found to explain  both  the baryon asymmetry and the dark matter of the Universe 
in a wide range of the couplings 
owing to the friction from the hidden gauge sector.
The experimental constraints require the ALP to have a mass between about $0.01$ and $30$~eV,
and additionally to be photophobic if the coupling to the electroweak anomaly is suppressed
by a scale below about $10^7$~GeV.  
\\

\noindent{\bf Acknowledgements}
\\ \\
The authors thank Alexandros Papageorgiou for helpful comments and discussion.  
This work was supported by IBS under the project code No.~IBS-R018-D1 (S.H.I.)
and  by the National Research Foundation  (NRF) of Korea grant funded 
by the Korea government: 
Grants No.~2018R1C1B6006061 and No.~2021R1A4A5031460 (K.S.J.  and Y.L.).
%


 \end{document}